\def \PRD {{\it Phys. Rev.} {\bf D}}

\def \JMP {{\it J. Math. Phys. }}

\def \bc {\begin{center}}
\def \ec {\end{center}}

\def \bfr {\begin{flushright}}
\def \efr {\end{flushright}}

\def \v {\vskip}
\def \ii {\'\i}

\def \ba {\begin{array}}
\def \ea {\end{array}}

\def \bea {\begin{eqnarray}}
\def \eea {\end{eqnarray}}

\def \be {\begin{equation}}
\def \ee {\end{equation}}

\def \L {{\cal L}}
\def \d {\hbox{d}\,}

\def \p {\partial}

\def \H {{\cal H}}
\def \x {{\bf x}}

\def \dirac {{\slash\!\!\!}} 

\documentstyle[12pt,a4]{article}

\begin{document}
\def \thesubsection {\Alph{subsection}}
\def \thesubsubsection {\thesubsection.\arabic{subsubsection}} 


\thispagestyle{empty}


\hfil{\bf Imperial-TP/95-96/24}\break

\begin{center}

{\bf TOWARD A FINITE-DIMENSIONAL FORMULATION 
OF QUANTUM FIELD THEORY}
\footnote[2]{This work is partially funded by the 
Spanish {\it Direcci\'on General 
de Ciencia y Tecnolog\ii a} (DGYCIT)}

 {\it Miguel Navarro}\footnote{http://www.ugr.es/$\,\widetilde{}$mnavarro;\ \ 
e-mail:mnavarro@ugr.es}
\end{center}
\v4mm
\noindent - The Blackett Laboratory, Imperial College, Prince 
Consort Road, London SW7 2BZ; United Kingdom.

\noindent - Instituto Carlos I de F\'\i sica Te\'orica y Computacional,
Facultad  de  Ciencias, Universidad de Granada, Campus de Fuentenueva,
18002, Granada, Spain.

\noindent  - Instituto de Matem\'aticas y F\'\i sica Fundamental, 
CSIC, Serrano 113-123, 28006 Madrid, Spain. 
\v5mm
\centerline{\bf Abstract}
\v2mm
\footnotesize 

Rules of quantization and equations of 
motion for a finite-dimensional 
formulation of Quantum Field Theory are proposed
which fulfill the following properties: 
a) both the rules of quantization and 
the equations of motion are covariant; 
b) the equations of evolution
are second order in derivatives and first order 
in derivatives of the space-time co-ordinates; and   
c) these rules of quantization and equations of motion 
lead to the usual (canonical) 
rules of quantization and the (Schr\"odinger) 
equation of motion of Quantum Mechanics in the particular 
case of mechanical systems. We also comment briefly on 
further steps to fully develop a satisfactory 
quantum field theory and the difficuties which 
may be encountered when doing so.     
\v3mm
\noindent PACS numbers: 03.65.Bz, \ 11.10.Ef, \  03.70.+k\hfil\break 
\normalsize
\newpage 
\setcounter{page}{1}
\section{Introduction}

At present the main goal of 
Theoretical Physics is to unify Quantum (Field) 
Theory and General Relativity. 
This task will probably require a previous 
reformulation of either of these theories or both of them. 

The standard way of quantizing a field theory 
-- and hence the actual form
of standard Quantum Field Theory (QFT) -- 
relies on the fact that 
Classical Field Theory (CFT) 
can be considered to be a generalization 
of Classical Mechanics (CM) in
which the finite number of 
degrees of freedom of the latter   
is replaced with an infinite (continuum) number in the former. 
In this formulation the fields are considered to be functions
$\varphi^a(\x)\,(t)\equiv\varphi^a_{\x}(t)$; 
that is, the spatial co-ordinates are regarded as labels 
(the discrete superindex $a$ labels the different 
fields in the theory). This description is supported 
primarily by the fact that it is a direct 
generalization of Quantum Mechanics (QM),  
which, as a theory with a vast range of predictions, 
is a source of great confidence. 
The standard framework requires, nonetheless, the use of 
functionals in place of 
ordinary functions as well as infinite-dimensional 
differential calculus,  which is plagued with 
ambiguities. These ambiguities are at the root 
of the renormalization problem. 

This fact also leads to a problem of foundation: 
if Classical Field Theory, which is based on a 
small number of ordinary functions over
the space-time, gives a description of the world, albeit rough and
primitive, why must Quantum Field Theory be
described with functionals -- that is, functions with an infinite
(continuum) number of arguments? 

All this raises the question of 
whether a description of the quantum theory of fields 
in terms of ordinary functions is possible or not.  

In fact, the kinematical description of a QFT of this type     
arises naturally from CFT  
provided that the latter, as a   
generalization of Classical Mechanics, is interpreted in a 
way different from the one  
that leads to the standard QFT \cite{[newrules]}. 
In this reading of CFT,   
all the co-ordinates of the space-time are
considered to be on the same footing,     
no special role is played by time.   
The fields are not taken to be an infinite (continuum) set of
functions of time but rather a discrete set of functions of all 
the space-time co-ordinates: 
$\varphi^a=\varphi^a(x),$ with $x=(\x, t)$
and $a$ a discrete label. 
Since there is a finite number of functions  
we shall refer to this approach
as finite-dimensional QFT as opposed to the standard or
infinite-dimensional QFT. 

The first steps towards a covariant finite-dimensional 
formulation of field theory were given by Born
\cite{[Born]}, Weyl \cite{[Weyl]}, de Donder \cite{[deDonder]} 
and  Carath\'{e}odory \cite{[Caratheodory]} as early as the 1930s. 
Further attempts are due to 
Good \cite{[Good1]} and Liotta \cite{[Liotta]}, and   
more recent ones to Tapia 
\cite{[Tapia]} and Kanatchikov \cite{[Kanatchikov]}. 
However, much of this effort has  
been focused on following routes to the quantum theory 
which closely mimick the one which, starting from 
Classical Mechanics, leads to the standard Quantum 
Mechanics. These routes pass, therefore, through developing  
a covariant canonical (Hamiltonian) 
formulation of the theory.  
The basic idea underlying this approach -- which can be 
referred to as the {\it bottom to top} approach -- is that, if a 
complete canonical formulation of the 
finite-dimensional description of Classical Field
Theory were found, the finite-dimensional 
quantum theory would then naturally follow. 

This procedure is legitimate, but it would perhaps be
more profitable to try to construct directly, 
by whatever the means, a self-consistent 
finite-dimensional covariant QFT. 
After all, Quantum Mechanics should arise as only a
limiting case of this QFT, and nothing guarantees 
that Poisson brackets, for instance, 
will play any role in the structure of the more general theory. 

This opposite route, which we might term {\it top to bottom}, has, in fact, 
been recently inaugurated with a proposal by Good of rules of 
quantization and equations of motion which give rise  
to a finite-dimensional QFT \cite{[Good]} .  
However, in the particular case of mechanical systems, 
Good's proposal does not lead to  
the standard Quantum Mechanics and,  
therefore, the resultant theory does not reproduce basic,  
long experimentally verified, predictions of 
standard QM \cite{[newrules]}. 
The proposal, as a whole, should therefore be discarded.    

In this context, the natural next step  toward  
a finite-dimen-\break sional QFT should be to find an 
alternative proposal which, while  
preserving the basic features of Good's framework,   
give rise to the standard QM in the case of 
mechanical systems, hence avoiding the experimental failure of 
Good's rules. The task therefore is to find rules 
of quantization and equations of motion such that: 

\begin{enumerate} 

\item Both rules of quantization and 
equations of motion must be explicitly covariant;  
i.e., space and time 
co-ordinates are treated on the same footing.   

\item Within the limits of mechanical systems these rules of 
quantization and equations of motion must 
reduce themselves to the familiar 
canonical rules of quantization and Schr\"odinger equation of 
evolution of ordinary Quantum Mechanics. 

\item The equations of evolution must be 
second order in derivatives and 
first order in derivatives of  
the space-time co-ordinates. 
\end{enumerate}

This task in fact constitutes   
the main goal of the present letter: to show, using   
an example to be presented below, that proposals which fulfill all the    
requirements above do exist. 

Although fully developing a quantum
theory is far beyond the scope of the present letter, some 
basic guidelines to carry on the present 
analysis are briefly commented in Section 3.  

\section{An improved proposal for equations of motion and  
rules of quantization}

Let us consider the Schr\"odinger equation of ordinary 
Quantum Mechanics:

\be i\frac{\d}{\d t}\Psi = \widehat{H}\Psi\label{1.1}\ee 
To generalize this equation to field theories, 
which are defined over a four-dimensional space-time manifold, 
we need a generalized Hamiltonian to be placed on its r.h.s., 
and a generalized ``time
derivative'' operator to be placed on its l.h.s.  
A generalizacion for the Hamiltonian is well known 
\cite{[Good],[newrules]}: the covariant Hamiltonian $\H$ which     
is obtained, from a Lagrangian $\L = \L(\phi^a,\p_\mu\phi^a)$,   
by means of the generalized covariant Legendre 
transform\footnote{If attention is paid to
other features of the Hamiltonian in Mechanics, its conservation
properties for instance, the energy momentum tensor 
${\Theta^\mu}_\nu={\phi^a}_\nu\,{\pi_a}^\mu-\delta^\mu_\nu\,{\cal L}$
may appear to be a more natural generalization in field theory. 
However, for the purposes of the present
letter, the covariant Hamiltonian $\H$ is equally good and allows us
to keep in line with Good's proposal. For the sake of being specific, 
we shall limit our present discussion to this case only.}: 

\be \H = \pi_a^\mu\p_\mu\phi^a -\L\>.\label{a1b}\ee
The covariant momenta $\pi^\mu_a$ are defined by: 

\be \pi^\mu_a= \frac{\p\>\L}{\p\>(\p_\mu\phi^a)}\label{a2}\ee

If we now write the Lagrangian in the 
following covariant Hamiltonian form

\be \L = \pi_a^\mu\p_\mu\phi^a - \H(\phi^a,\pi_a^\mu)\label{a3}\ee
its Lagrange equations of motion will also 
have a covariant Hamiltonian form: 

\bea \p_\mu\phi^a &=& \frac{\p\>\H}{\p\>\pi^\mu_a}\\
\p_\mu\pi^\mu_a&=& -\frac{\p\>\H}{\p\>\phi^a}\label{a4}\eea

With these ingredients, Good postulated the following 
quantum equation of motion \cite{[Good]} (see also \cite{[newrules]}): 

\noindent{\bf Good's quantum equation of motion}

\be -\frac{\p^2}{\p x^\nu \p x_\nu}\Psi(\varphi^a,x)\> 
= \widehat{\H}\Psi(\varphi^a,x)\label{goodevolution}\ee 
This equation of motion was supplemented with the following rules of
quantization: 
\v2mm 

\noindent{\bf Good's quantization rules}

\bea \varphi^a&\longrightarrow& \widehat{\varphi}^a
=\varphi^a\nonumber\\
\pi^\mu_a&\longrightarrow& \widehat\pi^\mu_a=
-\frac{\p^2}{\p \varphi^a\p x_\mu} \label{goodrules}\eea
This proposal, along with many attractive features, 
involves a number of undesired   
properties which prevent it from being a good starting point 
for a finite-dimensional formulation of QFT: 

a) The equation of motion is higher order in 
derivatives and (at least) second order in (space-)time derivatives.  

b) The proposal does not, in either 
the quantization rules or the evolution
equation, reproduce Quantum Mechanics in 
the particular case of a mechanical system. 

Either of these drawbacks is 
serious enough to rule out this proposal
as a good candidate for a finite-dimensional QFT. 
Moreover, it was shown in 
ref. \cite{[newrules]} that this theory 
leads to (measurable) predictions 
which do not agree with standard Quantum Mechanics. 
Hence, this proposal should be discarded. 

Fortunately, there are other proposals  
for quantization rules and evolution equations  
which are similar to Good's but behave much better. 

To motivate our proposal,  
let us consider the ordinary harmonic oscillator 
and the Dirac field. 
The respective Lagrangians can be written: 

\bea  \L_{HO} &=& a^*(i\dot{a} - a)\label{c1}\\
\L_D &=& \bar{\varphi}(i\dirac{\p}\varphi-\varphi)\label{c2}\eea
where $a$ ($a^*$) is the annihilation (creation) operator,    
$\dirac{\p}\equiv \gamma^\mu\p_\mu$, with $\gamma_\mu$ the Dirac's
matrices, and $\bar{\varphi}=\varphi^\dagger\gamma^0$.

Eqs. (\ref{c1}) and (\ref{c2}) tell us  
that the Dirac field is a higher-dimensional
generalization of the ordinary harmonic oscillator. The generalization
is accomplished by replacing the time derivative ${\d}/{\d t}$   
with the operator $\dirac{\p} =\gamma^\mu\p_\mu$. 

Mimicking that generalization, we can postulate the following 
equation of motion for our finite-dimensional QFT, which is intended 
to generalize ordinary QM\footnote{While the present letter was being 
refereed, Kanatchikov, independently and following different reasoning  
from ours, also arrived to this equation,  
and to other conclusions which are similar to ours  
\cite{[Kanatchikov2]}.} 

\v2mm
\noindent{\bf Dirac-like quantum equation of motion}

\be i\Gamma^\mu \p_\mu\Psi = \widehat{\H}\Psi
\label{neweqofmot}\ee 
Here $\Gamma_\mu$ are quantities which play a role similar 
to Dirac's matrices in the relativistic theory of the electron. 
However, we shall see later that further development of    
the theory may require the quantities $\Gamma_\mu$ not to be the 
Dirac matrices. For the moment, and for the sake of specificity, 
we may think of these as if they were the Dirac matrices, the Kemmer
matrices, or similar ones. 

The next step is to construct the operator $\widehat{\H}$;  
that is, we need quantization rules. Fortunately,  
once the quantities $\Gamma_\mu$ are at our disposal, it is
straightforward to propose quantization rules as well. These are: 
\v2mm
\noindent{\bf Dirac-like quantization rules} 

\bea \varphi^a&\longrightarrow& \widehat{\varphi}^a=\varphi^a\nonumber\\
\pi^\mu_a&\longrightarrow& \widehat\pi^\mu_a=
-i\Gamma^\mu\frac{\p}{\p\varphi^a} \label{newrules}\eea

The rules of quantization (\ref{newrules}) and the 
evolution equation (\ref{neweqofmot}) fulfill the three properties 
listed in the Introduction. In particular, 
unlike in Good's proposal,  
ordinary Quantum Mechanics is contained in this new proposal. 
Therefore, the vast amount of experimental predictions of
ordinary QM is entirely and automatically incorporated into our 
proposal. Hence, to rule out the new proposal 
we would have to look for a test which 
implied a genuine field system.  

Our proposal, as far as the quantum 
equation of motion is concerned,  
almost revives Good's proposal. 
In fact, if we identify the quantities $\Gamma_\mu$ with the Dirac
matrices and ``square'' the
equation of motion (\ref{neweqofmot}), we obtain an equation
similar to Good's, differing only in that 
the operator in its r.h.s. is not 
the covariant Hamiltonian $\widehat{\H}$,  
but rather its square. On the contrary, 
the quantization rules are sharply different and give
rise to Hamiltonian operators which also are strongly different.  

\section{Discussion and perspectives}

Our proposals for rules of quantization (\ref{newrules}) 
and equations of motion (\ref{neweqofmot}) 
fulfil the three properties in the Introduction. 
In this way, we improve Good's proposal in 
fundamental respects and actually solve 
the most serious objections which have 
been raised against it \cite{[newrules]}. 
 
A detailed analysis of the quantum theories that our proposal 
-- and related ones -- would lead to is a complex 
task which is beyond the scope of the present letter.
Let us, however, briefly comment on it. 

The next step should be to find 
a proposal with a natural, well-behaved, 
positive-definite scalar product $<\>|\>>$. In particular, 
and in analogy with QM, 
it seems natural to require that the scalar product 
of two wave functions $\Psi,\>\Phi$
should be space-time independent: 

\be   \p_\mu <\Psi|\Phi> = 0\label{pmucero}\ee 
This can be achieved if the equations of motion can be brought to the
form 

\be i\p_\mu \Psi=\widehat\H_\mu\Psi\label{ipmu}\ee
with $\H_\mu$ self-adjoint operators. 

Consider now that in our proposal we take  
$\Gamma_\mu$ to be the Dirac matrices $\gamma_\mu$.  
The natural ``scalar product'' is then: 

\be <\Psi|\Phi> 
= \int \d \varphi \bar{\Psi}\Phi, \quad 
\bar{\Psi}= \Psi^\dagger\gamma^0\label{sp}\ee 

However, neither is the norm $||\Psi||^2=<\Psi|\Psi>$ 
positive-definite nor is the product (\ref{sp})   
preserved by the space-time evolution. 

The first problem could be solved in a manner similar to 
the way in which the non-positivity of the Hamiltonian  
is solved when second quantizing the Dirac 
field -- by requiring that the wave function $\Psi$ be 
not a real field but rather a Grassmannian one. 
However, this solution raises the question of 
whether or not the resultant scalar product 
is independent on the representation of
the gamma matrices. 

That the scalar product is space-time dependent 
can be seen by considering wave functions $\Psi$  
for which $\widehat\H\Psi=\mu\Psi$, with 
$\mu\in\Re$. Then eq. (\ref{neweqofmot}) 
reduces to the Dirac equation, which have 
solutions in which the scalar product (\ref{sp}) 
is not space-time independent.  
This problem can be attributed to the fact that, for the case under
consideration, the 
operators $\widehat{\H}_\mu$ that appear on the r.h.s. 
of eq. (\ref{ipmu}) involve space-time derivatives. 
In fact, if we multiply eq. (\ref{neweqofmot}) 
by $\gamma_\mu$ and use the equality 

\be \gamma_\mu\gamma_\nu = g_{\mu\nu} -i\Sigma_{\mu\nu}\nonumber\ee 
we get 

\be i\p_\mu \Psi=\left(i\Sigma_{\mu\nu}\p^\nu+\gamma_\mu\widehat\H\right)
\equiv\widehat\H_\mu\Psi\ee 

Therefore, a satisfactory development of our proposal 
would require quantities $\Gamma_\mu$ which are not the Dirac
matrices. We should remark here that no reason has actually been put
forward to identify the quantities $\Gamma_\mu$ with 
the Dirac matrices. By now these quantities 
remain (almost) completely arbitrary; the only requirements 
are that the equations in which they are involved  
should be covariant under changes of reference frame. This can be
achieved with Dirac's matrices, but also with Kemmer's and others. 
The hope is that further development of our proposal will put stronger 
restrictions on those quantities and eventually determine them
completely. This would ultimately justify our introduction of them, 
which may appear to be a rather arbitrary ingredient of 
our proposal.  

A particularly interesting line of development would be to consider, rather
than the covariant Hamiltonian $\H$, 
the energy-momentum tensor $\Theta_{\mu\nu}$.
In this regard the developments in ref. \cite{[Tapia]}, 
where the classical dynamics of
fields is developed in terms of that quantity and generalized Poisson
brackets,  may be especially valuable. 

Finally, and for the sake of comparation,  
let us briefly show what the situation is in the standard QFT.  

In the pure Heisenberg picture of (standard) QFT,  
the momentum operators $P_\mu$ are such that 
 
\be e^{iP_\mu a^\mu}\widehat{O}_H(x) 
e^{-iP_\mu a^\mu}=\widehat{O}_H(x+a)\ee  
for any operator $\widehat{O}_H(x)$,  
in particular the fields $\widehat{\varphi}(x)$.  

By analogy with the Schr\"odinger picture of QM, 
let us remove {\it all} the space-time dependence from the operators 
and translate it into the wave functional by defining: 

\bea \widehat{O}_S&=& e^{-iP_\mu x^\mu}
\widehat{O}_H (x) e^{iP_\mu x^\mu}\nonumber\\
\Psi(x)_S&=&e^{iP_\mu x^\mu} \Psi_H\label{si}
\eea 
The wave functionals $\Psi(x)_S$ now obeys   
generalized Schr\"odinger equations  (\ref{ipmu}): 

\be i\p_\mu \Psi(x)_S= P_\mu \Psi(x)_S
\label{ipmu2}\ee 

In this way, we have recovered 
a picture of the standard QFT (which can be called {\it
pure Schr\"odinger} picture of QFT),  
which incorporates much of 
the spirit of the finite-dimensional QFT,  
although not the basic requirement -- 
that the wave functions must be ordinary functions. 
On the contrary, the functional $\Psi$ in eqs. 
(\ref{si},\ref{ipmu2}) are defined, not
over the space of fields $\varphi^a$, 
but over the {\it phase space}
of the theory which, 
in general, is infinite-dimensional \cite{[cps]}.  \\

\v1mm 

\noindent{\bf Acknowledgements.} 
The author is grateful to the Imperial 
College, where this paper was written, for its hospitality,  
and to the referee for his/her very useful comments and suggestions.  
The author also thanks the Spanish MEC, CSIC and IMAFF (Madrid) 
for a research contract.


\begin{thebibliography}{99}


\bibitem{[Born]}M. Born, {\it Proc. Roy. Soc. London} {\bf A} 
{\bf 143}, 410 (1934).

\bibitem{[Weyl]}H. Weyl, {\it Phys. Rev.} {\bf 46}, 505 (1934).

\bibitem{[deDonder]}Th. de Donder, {\it Theorie invariantive du calcul des 
variations} (Gauthier--Villars, Paris, 1935).

\bibitem{[Caratheodory]} C. Carath\'{e}odory, {\it Acta Sci. 
Math.} (Szeged) 4 (1929) 193

\bibitem{[Good1]}R. H. Good, {\it Phys. Rev.} {\bf 93}, 239 (1954).

\bibitem{[Liotta]}R. S. Liotta, {\it Nuovo Cimento} {\bf 3}, 438 (1956).

\bibitem{[Tapia]}V. Tapia, {\it Nuovo Cimento} {\bf B} {\bf 102}, 123 (1988).

\bibitem{[Kanatchikov]} Igor V. Kanatchikov, ``On the Canonical 
Structure of the De Donder-Weyl Covariant Hamiltonian Formulation of 
Field Theory. I. Graded Poisson brackets and equations of motions'', 
PITA 93/41; hep-th/9312162.

\bibitem{[Good]}R. H. Good, {\it J. Math. Phys} {\bf 35}, 
3333 (1994); {\it ibid} {\bf 36}, 707 (1995).

\bibitem{[newrules]} M. Navarro, \JMP{\bf 36} (1995)6665. 

\bibitem{[cps]} For a covariant characterization 
of the phase space of
a system as the space of all its (classical) trajectories 
see refs. C. Crnkovi\' c and E. Witten, 
in {\it Three Hundred Years of Gravitation},
eds. S. W. Hawking and W. Israel (Cambridge, 1987), p. 676; 
A. Ashtekar, L. Bombelli and O Reula, 
in {\it Mechanics, Analysis
and Geometry: 200 Years after Lagrange}, 
ed. M. Francavigilia
(ESP, 1991), p. 417; and also  
J. Navarro-Salas, M. Navarro and C.F. Talavera, 
 \PRD{\bf 52} (1995)6831-6839. 

\bibitem{[Kanatchikov2]} I.V. Kanatchikov, ``Towards the Born-Weyl
Quantization of Fields'', to appear in {\it Int. J. Th. Phys.}, 
quant-ph/9712058. 

\end{thebibliography}
\end{document}